\documentclass[12pt]{article}
\begin{document}

\begin{center}

{\bf\large CURVATURE COUPLING AND ACCELERATED}

\vspace{3mm}

{\bf\large EXPANSION OF THE UNIVERSE}

\vspace{6mm}

{\bf Alexander B. Balakin}

\vspace{5mm}

Department of General Relativity and Gravitation

Kazan State University, 420008 Kazan, Russia

{\it E-mail: Alexander.Balakin@ksu.ru}

\end{center}

\begin{abstract}
A new exactly solvable model for the evolution of relativistic
kinetic system interacting with an internal stochastic reservoir
under the influence of a gravitational background expansion is
established. This model of self-interaction is based on the
relativistic kinetic equation for the distribution function
defined in the extended phase space. The supplementary degree of
freedom is described by the scalar stochastic variable (Langevin
source), which is considered to be the constructive element of the
effective one-particle force. The expansion of the Universe is
shown to be accelerated for the suitable choice of the non-minimal
self-interaction force.
\end{abstract}


\section{Introduction}
\label{introduction}

Observations of supernovae at high redshift show that the
expansion of the Universe is accelerating. To explain this fact
in the framework of Friedmann - Lema$\hat{{\rm i}}$tre -
Robertson - Walker (FLRW) model, one has to require that a cosmic
medium is characterized by negative pressure \cite{RT,Riess}. One
of the possible explanations is the existence of dark energy
(see, e.g., the reviews \cite{aastar1,Ellis}). There exists a
number of dark energy candidates, the best known are a
cosmological constant and different quintessence scenarios
(\cite{RaPeeb,caldwell,amendola,ZP,Matar,Uzan,Capo,Frolov}).
Negative pressure may also be the consequence of self-interactions
in gas models of the Universe \cite{ZiBa01,ZSBP,ZBEntr,S,NJP}. In
particular, an ``antifrictional'' force, self-consistently
exerted on the particles of the cosmic substratum, was shown to
provide an alternative explanation for an accelerated expansion
of the Universe \cite{ZSBP,S}. This approach relies on the fact
that the cosmological principle is compatible with the existence
of a certain class of (hypothetical) microscopic one-particle
forces, which manifest themselves as ``source'' terms in the
macroscopic perfect fluid balance equations. These sources can be
mapped on an effective negative pressure of the cosmic medium.

It is worth pointing that effective self-interaction forces can
be regarded as a specific non-minimal coupling of the cosmic gas
to the Ricci scalar, Ricci tensor and Riemann tensor. Generally,
a force which explicitly depends on curvature quantities
describes a coupling of matter with the space-time curvature
which goes beyond Einstein's theory. However, mapping the
non-minimal interaction on an imperfect fluid degree of freedom
admits a self-consistent treatment on the basis of general
relativity. This may be considered as a gas dynamical counterpart
to the non-minimal couplings of scalar fields or those of
higher-order gravity theories. A (non-minimal) fluid interaction
is designed so that it results in the desired cosmic evolution.
Designing the coupling for description of a specific dynamics has
already been used earlier for interacting two-component models
\cite{ZP}. Here this idea is applied to the case of a
one-component fluid, which is self-consistently coupled with the
Riemann tensor. As a characteristic feature of this approach,
Hubble rate and deceleration parameter explicitly enter the
microscopic dynamics, giving rise to a self-consistent coupling
of the latter to the gravitational field equations.

The paper is organized as follows. In section 2 the formalism of
one-particle distribution function in the extended phase space
associated with a homogeneous and isotropic, spatially flat
Universe is established. In section 3 macroscopic properties of
the ultrarelativistic kinetic system are discussed, an effective
stress-energy tensor is obtained, and the conditions of
accelerated expansion of the Universe for power-law and
exponential scenarios are investigated in detail. A brief summary
is given in section 4.

\section{Formalism of One-Particle Distribution \\ Function in the Extended Phase Space}

\subsection{Kinetic Equation}

The idea of the phase space extension, based on the covariant
formalism of Cartan's differentiation and integration,  was
proposed in \cite{BLL,AAV}. Along a line of development of the
relativistic kinetic theory (see, \cite{Ehl,Stew,Groot} and
references therein) a numerous applications of that formalism
have been elaborated by different authors (see, e.g.,
\cite{RH,WI,FT,EH,LI}).

The simplest kinetic system with supplementary degree of freedom
can be described in terms of the 8+1 -dimensional scalar
distribution function
\begin{equation}
\Phi = \Phi \left( x^{i}, p^{k} \,, \omega \right), \label{1}
\end{equation}
depending on coordinates $x^{i} $,  particle momentum four-vector
$p^k$, and on the random scalar variable $\omega$, which is
called random Langevin's source. The distribution function
satisfies the kinetic equation
\begin{equation}
\frac{p^{i}}{m c} \left( \frac{\partial{}}{\partial{x^{i}}} -
\Gamma ^{k}_{il} p^{l} \frac{\partial{}}{\partial{p^{k}}} \right)
\Phi  + \frac{\partial{}}{\partial{p^{i}}} \left(F^{i} \Phi
\right) + \frac{\partial{}}{\partial{\omega}} \left( {\cal H}
\Phi \right)  = 0 \,, \label{2}
\end{equation}
where $\Gamma^{k}_{il}$ are the Christoffel symbols, associated
with the background metric $g _{ik}$. Characteristics equations,
corresponding to (2), form three subsystems:
\begin{equation}
\frac{d p ^{i}}{ds} + \Gamma ^{i} _{kl} p ^{k} \frac{d x ^{l}}{d
s}= F ^{i} \,, \quad \frac{d \omega}{ds} = {\cal H} \,, \quad
\frac{d x ^{i}}{d s} = \frac{p ^{i}}{m c} \,. \label{3}
\end{equation}
The first equation in (3) is the well-known equation of general
relativistic particle dynamics under the influence of the force
$F^{i}\left( x^{i}, p^{k} \,, \omega \right)$. The second
equation is the evolutionary equation for the scalar random
variable $\omega$, which is modeling the stochastic influence of
the environment on the particle. Using (3), one can write the rate
of evolution of the distribution function $\Phi$ in the form
\begin{equation}
\frac{d \Phi}{d s} = - \left[ \frac{\partial{F ^{i}}}{\partial{p
^{i}}}  + \frac{\partial{{\cal H}}}{\partial{\omega}} \right]
\Phi \,. \label{4}
\end{equation}
The moments of the distribution function (1) can be obtained  by
averaging over five-dimensional statistical ensemble:
\begin{equation}
N^{i} (x) \equiv \int {d P  d \omega \cdot \Phi \cdot p^{i}} \,,
\quad T^{ik} (x) = \int {d P d \omega \cdot \Phi \cdot p ^{i} p
^{k}} \,. \label{5}
\end{equation}
Denoting the term $\int d \omega \cdot \Phi $ by  $ f (x, p)$, one
obtains from (5) the standard formulas for the particle number
density vector $N^{i} (x)$ and for the stress - energy tensor
$T^{ik} (x)$. The standard definition for the entropy flux vector:
\begin{equation} S ^{i} (x) \equiv  - k _{B} c
\int {d P d \omega \cdot \Phi \cdot p ^{i}  \left[ \ln h^3 \Phi -
1 \right]} \label{6}
\end{equation}
is used. The transport equations for the particle number,
stress-energy and for the entropy have the form, respectively:
\begin{equation}
\nabla_{i} N ^{i}(x) = 0 \,, \quad \nabla_{k} T^{ik}(x) = mc \int
{d P d \omega \cdot \Phi \cdot F^i} \,, \label{7}
\end{equation}
\begin{equation}
\sigma (x) \equiv  \nabla _{i} S ^{i} = k _{B} m c ^{2} \int {d P
d \omega \cdot \Phi \left[ \frac{\partial{F ^{i}}}{\partial{p
^{i}}}  + \frac{\partial{{\cal H}}}{\partial{\omega}} \right]}
\,. \label{8}
\end{equation}

\subsection{Cosmological Background}

The spatially flat FLRW solution of Einstein's equations is
considered:
\begin{equation}
ds^{2} =  c^2 d t^2 - a^{2}(t) \left[(dx^{1})^{2} + (dx^{2})^{2} +
(dx^{3})^{2} \right] \,, \label{9}
\end{equation}
\begin{equation}
\ddot{a} = - \frac{4\pi G}{3} a \left( {\cal E} + 3 {\cal P}
\right), \quad \dot{a}^2 = \frac{8\pi G}{3} a^2 {\cal E} \,.
\label{10}
\end{equation}
Here ${\cal E}$ and ${\cal P}$ are the effective energy density
and effective pressure, respectively. These quantities are known
to be the eigenvalues of the conserved effective stress-energy
tensor
\begin{equation}
T^{ik}_{({\rm eff})} = ( {\cal E}  +  {\cal P} ) U^i U^k - g^{ik}
{\cal P} \,,  \quad  \nabla_k T^{ik}_{({\rm eff})} = 0 \,.
\label{11}
\end{equation}
The velocity four-vector is equal to $U^i = \delta^i_0$, as usual.

\subsection{Force Field Modeling}

The motion of Brownian particle in the framework of classical
dynamics can be modeled by the force three-vector \cite{Balescu}
\begin{equation}
\vec{{\cal F}} = - \lambda (\vec{v} - \vec{U}) + \kappa \vec{\xi}
\,. \label{12}
\end{equation}
The first term of this force is the Stokes friction force, it
vanishes when the particle velocity three-vector $\vec{v}$
coincides with the medium flow three-vector $\vec{U}$. The second
term is the stochastic Langevin force with random three-vector
$\vec{\xi}$. The covariant generalization of the Stokes and the
Langevin forces is well-known:
\begin{equation}
F^i_{{\rm Stokes}} =  \lambda \left(\delta^i_k - \frac{p^i
p_k}{(p^l p_l)} \right) U^k \,, \quad F^i_{{\rm Langevin}} =
\kappa \left(\delta^i_k - \frac{p^i p_k}{(p^l p_l)} \right) \xi^k
\,. \label{13}
\end{equation}
The projector with respect to particle four-momentum in the
parentheses provides the forces to be orthogonal to $p^i$: $F^i
p_i = 0$.

In this paper we consider the force four-vector
\begin{equation}
F^i( x^{i}, p^{k}, \omega ) = \omega \frac{q} {mc^2} \left[(p^l
p_l) \delta^i_k  - p^i p_k \right] U^k \,, \label{14}
\end{equation}
where $q$ is considered to be a function of cosmological time and
is a subject of modeling. It was shown in \cite{98} that the
existence of the force with such a structure is compatible with
the symmetry requirement
\begin{equation}
\pounds_{\zeta} g_{ik} = 2 \Psi \left(g_{ik} - \frac{\zeta_i
\zeta_k}{(\zeta^l \zeta_l)}  \right) \,, \label{15}
\end{equation}
formulated in terms of Lie derivative. The corresponding time-like
vector $\zeta^i$ is shown to exist for FLRW space-time, and the
explicit example of the so-called generalized equilibrium states
has been found in \cite{98}. Starting from this fact we have
studied in \cite{ZSBP,NJP} the consequences of the appearance of
the antifriction force. The force (14) gives the Stokes force
when $\omega q m = \lambda$ and the Langevin force when $\omega q
m U^i = \kappa \xi^i$.

The function ${\cal H}$ is modeled as follows:
\begin{equation}
{\cal H}( x^{i}, p^{k}, \omega ) = \omega \frac{\chi}{mc^2} (p^k
U_k) \,. \label{16}
\end{equation}
The coefficients $q$ and $\chi$ are considered to be scalar
functions, depending on the Ricci scalar $R$, on the Hubble
parameter $H$ (which is proportional to the scalar of expansion
$H = \frac{1}{3} \nabla_k U^k$) and on the scalar $\hat{R} \equiv
R_{ik} U^i U^k$. The terms $F^i$ and ${\cal H}$ are linear in
dimensionless random variable $\omega$, i.e., one can indicate a
stochastic Langevin's source $\omega$ as multiplicative one
\cite{mult}. The function ${\cal H}$ in the form (16) guarantees
that the trivial value $\omega \equiv 0$ is a singular solution
of the dynamic equation (3). Since
\begin{equation}
\frac{\partial{ F^{i}}}{\partial{p^{i}}} + \frac{\partial{ {\cal
H}}}{\partial{\omega}} = \frac{(p^k U_k)}{mc^2} (\chi - 3 q
\omega) \neq 0 \,, \label{17}
\end{equation}
the entropy production scalar $\sigma$ (8) does not vanish.

\subsection{Solution to the Characteristics Equations}

Supposing that in the FLRW space-time (9) all the macroscopic
functions depend on time only, and using the consequence of the
characteristics equations
\begin{equation}
\frac{cdt}{d s} = \frac{p^0}{mc} \,, \label{18}
\end{equation}
one can rewrite the dynamic equations in terms of $t$
\begin{equation}
\dot{p}_\alpha = - q \omega  p_\alpha  , \quad \dot{\omega} =
\chi \omega \,. \label{19}
\end{equation}
The dot denotes the derivative with respect to time, and $\alpha =
1,2,3$. The solutions of (19) are the following:
\begin{equation}
p_{\alpha} = C_{\alpha} e^{- \Omega J(t, t_0)}, \quad J(t, t_0)
\equiv \int^t_{t_0} dt q(t) I(t, t_0) \,, \label{20}
\end{equation}
\begin{equation}
\omega = \Omega \cdot I(t, t_0) \,, \quad I(t, t_0) \equiv
e^{\int^t_{t_0} dt \chi(t)} \,. \label{21}
\end{equation}
By definition $J(t_0, t_0) = 0$ and  $I(t_0, t_0) = 1$, providing
the relations $p_\alpha(t_0) = C_\alpha$ and $\omega(t_0) =
\Omega$. Using the solution for $p_\alpha$ and the normalization
condition $p^i p_i = m^2 c^2$, one can obtain the $p^0 = p_0$
component of the particle momentum:
\begin{equation}
p^0(t) = \sqrt{m^2 c^2 + a^{-2}(t) {\bf C}^2  e^{-2 \Omega
J(t,t_0)}} \,. \label{22}
\end{equation}
Here ${\bf C}^2 \equiv C_1^2 + C_2^2 + C_3^2$, and the
distribution function $\Phi$ takes the form
\begin{equation}
\Phi = \Phi_0 ({\bf C}^2, \Omega, t_0) e^{3 \Omega J(t,t_0)}
I^{-1}(t,t_0) \delta(\sqrt{(p,p) - m^2c^2)} \,. \label{23}
\end{equation}
The transformation of the generalized volume element gives the
formula
\begin{equation}
\int{ d P d \omega \cdot \Phi \left\{ \cdot \cdot \cdot \right\}}=
\int {\frac{dC_{1} dC_{2} dC _{3}}{a^3(t) \sqrt{m^2 c^2 +
a^{-2}(t) {\bf C}^2  e^{-2 \Omega J(t,t_0)}}} d \Omega \Phi_{0}
\left\{\cdot \cdot \cdot \right\}} \,. \label{24}
\end{equation}

\section{Macroscopic Properties of the Kinetic System}

\subsection{The Structure of Macroscopic Moments}

Particle number is a conserved quantity, thus, one obtains
\begin{equation}
N^0 a^3(t) = const , \quad N^i(t) = N(t_0)
\left[\frac{a(t_0)}{a(t)}\right]^3 \delta^i_0 \,. \label{25}
\end{equation}
The stress-energy tensor takes the form
\begin{equation}
T^i_{k}(t) = \int \frac{d^3 C  \cdot d\Omega}{ a^3(t) \sqrt{m^2
c^2 + a^{-2}(t) {\bf C}^2  e^{-2 \Omega J(t,t_0)}}} \Phi_0({\bf
C}^2, \Omega, t_0) \cdot  \tau^i_{k}({\bf C}^2, J(t,t_0)) \,,
\label{26}
\end{equation}
where
\begin{equation}
\tau^i_{k}({\bf C}^2, J(t,t_0)) \equiv a^{-2} {\bf C}^2  e^{-2
\Omega J} diag \left\{ \left[m c a {\bf |C|}^{-1}  e^{ \Omega J}
\right]^2 + 1 , \ - \frac{1}{3} , \ - \frac{1}{3} , \ -
\frac{1}{3} \right\} \,. \label{27}
\end{equation}
In general case the stress-energy tensor can not be represented in
the analytic form, thus, for the sake of simplicity the
ultrarelativistic limit is used to illustrate the idea. If the
term, containing  $m^2c^2$ in the function (27), is negligible in
comparison with the second term, and if the initial distribution
function $\Phi_0$ is multiplicative, i.e., $\Phi_0({\bf C}^2,
\Omega, t_0) = f_0 ({\bf C}^2, t_0) \cdot \Psi(\Omega,t_0)$, one
can obtain the stress - energy tensor in the form:
\begin{equation}
T^i_{k}(t) =  A(t,t_0) W(t_0) \left[\frac{a(t_0)}{a(t)}\right]^4
diag \left\{ 1, \ - \frac{1}{3}, \ - \frac{1}{3}, \ -
\frac{1}{3}\right\} \,. \label{28}
\end{equation}
Here
\begin{equation}
A(t,t_0) \equiv   \int d\Omega \Psi_0(\Omega, t_0) e^{- \Omega
J(t,t_0)} \equiv \langle e^{- \Omega J} \rangle  \label{29}
\end{equation}
is a statistical factor. The energy density scalar
\begin{equation}
W(t_0) \equiv   \int \frac{d^3 C}{ a^4(t_0)} \vert {\bf C} \vert
f_0({\bf C}^2, t_0)  \label{30}
\end{equation}
is known to be equal to
\begin{equation}
W_{{\rm bosons}}(t_0) = \frac{8 k_B^4 \pi^5 }{15 h^3 c^3} T^4(t_0)
\,, \quad W_{{\rm fermions}}(t_0) = \frac{7 k_B^4 \pi^5 }{30 h^3
c^3} T^4(t_0) \label{31}
\end{equation}
for ultrarelativistic (massless) bosons and for ultrarelativistic
(massless) fermions. The entropy production scalar
\begin{equation}
\sigma(t) = k_B c  N(t) \int d\Omega \Psi_0(\Omega,t_0) (\chi - 3
q \Omega I)  \label{32}
\end{equation}
happens to be proportional to particle density scalar $N(t)$.

\subsection{Properties of the Averaged Macroscopic Moments}

Let us suppose that the distribution over $\Omega$ can be
described by the Gaussian function
\begin{equation}
\Psi_0(\Omega,t_0) = \frac{1}{\sqrt{\pi} D}
e^{-\frac{\Omega^2}{D^2}} \,, \label{33}
\end{equation}
where $D$ is a dispersion parameter of this distribution at the
moment $t_0$. For this case the direct calculations give
\begin{equation}
\sigma(t) = k_B c \chi  N(t)   \label{34}
\end{equation}
and
\begin{equation}
A(t,t_0) =  e^{ \frac{1}{4} D^2 J^2(t,t_0)}  \,. \label{35}
\end{equation}
Thus, the properties of the macroscopic moments of the
distribution function are predetermined by the properties of the
functions $J(t,t_0)$ (20) and $I(t,t_0)$ (21).

\subsection{Effective Stress-Energy Tensor}

The relation (7) demonstrates that the energy and the momentum of
the kinetic system do not conserve, since the balance equation
has the source term in the right-hand side. Calculating this
source term using the expression (14) for the force $F^i$ one
obtains
\begin{equation}
\nabla_k T^{ik} = q I(t,t_0) \int \frac{d^3C \Omega d \Omega
\Phi_{0}}{c a^3(t) \sqrt{m^2 c^2 + a^{-2}(t) {\bf C}^2  e^{-2
\Omega J}}} [m^2 c^2 U^i - p^i (U_k p^k)] \,. \label{36}
\end{equation}
For the ultrarelativistic model with multiplicative distribution
function this expression is simplified:
\begin{equation}
\nabla_k T^{ik} = - \delta^i_0 q I(t,t_0)
\left[\frac{a(t_0)}{a(t)} \right]^4 \int \Omega d \Omega
\Psi_0(\Omega, t_0) e^{- \Omega J} \int \frac{d^3C}{c a^4(t_0)}
\vert {\bf C} \vert f_0({\bf C}^2, t_0) \,. \label{37}
\end{equation}
For the Gaussian function (33) the integrals in the balance
equation take the form
\begin{equation}
\nabla_k T^{ik} = \delta^i_0 q I(t,t_0) W(t_0)
\left[\frac{a(t_0)}{a(t)} \right]^4  \frac{D^2}{2c} J(t,t_0) e^{
\frac{D^2}{4} J^2(t,t_0)}\,. \label{38}
\end{equation}
For $i=1,2,3$ these equations are trivial, the only informative
equation is the one for $i=0$. This equation (the so-called
scalar balance equation) may be written as follows:
\begin{equation}
\dot{W}(t) + 3 H [W(t) + P(t)] =  q I(t,t_0) W(t_0)
\left[\frac{a(t_0)}{a(t)} \right]^4  \frac{D^2}{2} J(t,t_0) e^{
\frac{D^2}{4} J^2(t,t_0)}\,, \label{39}
\end{equation}
where $W(t) = T^{00}$. The equation (39) takes the form of
conservation law
\begin{equation}
\dot{W} + 3 H (W + {\cal P}) = 0 \,, \label{40}
\end{equation}
if we introduce an effective pressure ${\cal P} \equiv P + \Pi$,
where
\begin{equation}
\Pi \equiv -   q I(t,t_0) W(t_0) \left[\frac{a(t_0)}{a(t)}
\right]^4 \frac{D^2}{6 H(t)} J(t,t_0) e^{ \frac{D^2}{4}
J^2(t,t_0)}  \label{41}
\end{equation}
is considered to be non-Pascal pressure. The reconstruction of
the effective stress-energy tensor (11) is possible if we put
\begin{equation}
{\cal E} = W(t) = W(t_0) \left[\frac{a(t_0)}{a(t)} \right]^4
e^{\frac{D^2}{4} J^2} \,, \label{42}
\end{equation}
\begin{equation}
{\cal P} =  \frac{1}{3} W(t_0) \left[\frac{a(t_0)}{a(t)} \right]^4
e^{ \frac{D^2}{4} J^2} \left[ 1 - q I(t,t_0) \frac{D^2}{2 H(t)}
J(t,t_0) \right] \,. \label{43}
\end{equation}
The formulas (42) and (43) manifest the following feature. The
coefficient $\left[\frac{a(t_0)}{a(t)} \right]^4$ describes the
decreasing of the average energy and pressure, which corresponds
to the standard law of evolution of ultrarelatistic FLRW model.
The coefficient $e^{\frac{D^2}{4} J^2}$ describes the increase of
the mentioned quantities due to the gas (fluid) stochastic
self-interaction. The interplay between these two processes could
clarify the question whether ${\cal E}$ is the increasing or
decreasing quantity. The sign of ${\cal P}$ can be both positive
or negative during the different time intervals. The expansion of
the Universe happens to be accelerated ($\ddot{a} > 0$)(see,
(10)), when
\begin{equation}
{\cal E} + 3 {\cal P} = 2 W + 3 \Pi < 0 \,. \label{44}
\end{equation}
It is possible when
\begin{equation}
q(t) I(t,t_0) \frac{D^2}{4 H(t)} J(t,t_0) > 1 \,. \label{45}
\end{equation}
Let us discuss  this inequality in detail.

\subsection{Modeling of the $q$ and $\chi$ Functions}

The functions $q$ and $\chi$  can be modeled using $R$, $\hat{R}$
and $H$ functions:
\begin{equation}
R = - 6 \left[\frac{\ddot{a}}{a} + \left(\frac{\dot{a}}{a}
\right)^2 \right] = - 6 (\dot{H} + 2 H^2) \,, \label{46}
\end{equation}
\begin{equation}
\hat{R} = R_{00} = - 3 \frac{\ddot{a}}{a}  = -3  (\dot{H} +  H^2)
\,, \quad H \equiv \frac{\dot{a}}{a} \,.\label{47}
\end{equation}
Taking into account these relations,  $q$ and $\chi$ can be
rewritten as a functions of $H$ and $\dot{H}$. Since $q$ and
$\chi$ have a dimension of the inverse time, it is proposed to use
the following simplest structures:
\begin{equation}
\chi(t) = \chi_0 H(t) + \chi_1 \frac{\dot{H}}{H} \,, \quad q(t) =
q_0 H(t) + q_1 \frac{\dot{H}}{H} \,. \label{48}
\end{equation}
Here $\chi_0$, $\chi_1$, $q_0$ and $q_1$ are dimensionless
constants. From (21) one obtains explicitly
\begin{equation}
I(t,t_0) = \left[\frac{a(t)}{a(t_0)} \right]^{\chi_0} \cdot \left[
\frac{H(t)}{H(t_0)} \right]^{\chi_1}  \,. \label{49}
\end{equation}
The function $J(t,t_0)$
\begin{equation}
J(t,t_0) = \int^t_{t_0} dt \left[\frac{a(t)}{a(t_0)}
\right]^{\chi_0} \cdot \left[ \frac{H(t)}{H(t_0)}
\right]^{\chi_1} \left[  q_0 H(t) + q_1 \frac{\dot{H}}{H} \right]
\label{50}
\end{equation}
can be calculated explicitly for two well-known cases. These
results are presented briefly in the following subsections 3.4.1.
and 3.4.2.

\subsubsection{Power-Law Expansion}

When
\begin{equation}
\frac{a(t)}{a(t_0)} = \left(\frac{t}{t_0}\right)^{\gamma} \,,
\quad H(t) = \frac{\gamma}{t} \,, \quad \dot{H} = -
\frac{\gamma}{t^2} \,, \label{51}
\end{equation}
we obtain immediately the following functions:
\begin{equation}
J(t,t_0) =  \frac{(\gamma q_0 - q_1)}{(\gamma \chi_0 - \chi_1)}
\cdot  \left[I(t,t_0) - 1 \right] \,, \quad I(t,t_0) =
\left(\frac{t}{t_0}\right)^{\gamma \chi_0 - \chi_1}  \,.
\label{52}
\end{equation}
The predictions concerning the accelerated expansion depend on
the sign of the parameter $\gamma \chi_0 - \chi_1$. Let us
consider three standard cases.

\noindent
{\it First case:} $\gamma \chi_0 > \chi_1$.

\noindent
The inequality (45) is satisfied when
\begin{equation}
\left(\frac{t}{t_0} \right)^{\gamma \chi_0 - \chi_1} >
\left[\frac{1}{2} + \sqrt{\frac{1}{4} + \frac{4 \gamma (\gamma
\chi_0 - \chi_1) }{D^2 (\gamma q_0 - q_1)^2}} \right] \,.
\label{53}
\end{equation}

\noindent {\it Second case:} $\gamma \chi_0 < \chi_1$.

\noindent When the discriminant is positive, i.e., $ 16 \gamma
|\gamma \chi_0 - \chi_1 | < D^2 (\gamma q_0 - q_1)^2 $, the
inequality (45) takes place, if
\begin{equation}
\left[\frac{1}{2} - \sqrt{\frac{1}{4} - \frac{4 \gamma |\gamma
\chi_0 - \chi_1| }{D^2 (\gamma q_0 - q_1)^2}} \right] <
\left(\frac{t}{t_0}\right)^{- |\gamma \chi_0 - \chi_1|} <
\left[\frac{1}{2} + \sqrt{\frac{1}{4} - \frac{4 \gamma |\gamma
\chi_0 - \chi_1| }{D^2 (\gamma q_0 - q_1)^2}} \right]
 \,. \label{54}
\end{equation}

\noindent
{\it Third case:} $\gamma \chi_0 = \chi_1$.

\noindent
For this special case the formulas (52) give
\begin{equation}
I(t,t_0) = 1 \,, \quad J(t,t_0) =  (\gamma q_0 - q_1) \cdot
\log{\left(\frac{t}{t_0}\right)} \,. \label{55}
\end{equation}
The inequality (45) is satisfied when
\begin{equation}
\log \left(\frac{t}{t_0}\right) >  \frac{4 \gamma}{(\gamma q_0
 - q_1)^2 D^2} \,. \label{56}
\end{equation}

\subsubsection{Exponential Expansion}

When
\begin{equation}
\frac{a(t)}{a(t_0)} = e^{H_0 (t - t_0)} \,, \quad H(t) = H(t_0) =
H_0 \,, \quad \dot{H} = 0 \,, \label{57}
\end{equation}
the integration gives
\begin{equation}
J(t,t_0)  = \frac{q_0}{\chi_0} \left[ I(t,t_0) - 1 \right] \,,
\quad I(t,t_0) = e^{\chi_0 H_0 (t - t_0)} \,. \label{58}
\end{equation}

\noindent
{\it First case:} $\chi_0 > 0$.

\noindent
The inequality (45) is satisfied when
\begin{equation}
(t - t_0) >  \frac{1}{\chi_0 H_0} \log\left[\frac{1}{2} +
\sqrt{\frac{1}{4} + \frac{4 \chi_0}{D^2 q_0^2}} \right] \,.
\label{59}
\end{equation}

\noindent {\it Second case:} $\chi_0 < 0$, $|\chi_0| < \frac{D^2
q_0^2}{16}$.

\noindent
The inequality is satisfied when
\begin{equation}
\left[\frac{1}{2} - \sqrt{\frac{1}{4} - \frac{4 |\chi_0|}{D^2
q_0^2}}\right] < e^{- |\chi_0| H_0 (t - t_0)} < \left[\frac{1}{2}
+ \sqrt{\frac{1}{4} - \frac{4 |\chi_0|}{D^2 q_0^2}} \right] \,.
\label{60}
\end{equation}

\noindent
{\it Third case:} $ \chi_0 = 0$

\noindent
The formulas (58) give
\begin{equation}
I(t,t_0) = 1 \,, \quad J(t,t_0) =   q_0 H_0 (t - t_0)  \,.
\label{61}
\end{equation}
The inequality (45) is satisfied when
\begin{equation}
(t - t_0) >  \frac{4}{q_0^2 H_0 D^2}  \,. \label{62}
\end{equation}

\section{Discussion}

Here the model of evolution of the self-interacting gas (fluid)
Universe is presented. The model is based on the suggestion that
the expansion of the Universe gives rise to the specific
non-equilibrium self-interaction in the kinetic system. The force
effecting the particle can be classified as a Stokes friction
force, since it disappears when particle co-moves with the system
as a whole. The force under consideration can be indicated as a
Langevin force, since it is proportional to the random scalar
variable, and the particle motion can be classified as a sort of
Brownian motion. Finally, this force can be called tidal or
curvature induced force, since it depends on Hubble rate and its
derivative, which are known to form the Riemann tensor, Ricci
tensor, and Ricci scalar. In this sense one can say that we deal
with curvature induced Stokes-Langevin force.

In the subsection 3.4. the time intervals are found, during which
the Universe expands with acceleration. The explanation of such a
behaviour may be the following. The self-interaction in the gas
(fluid) produces the growth of the averaged energy and pressure
in the system, in contrast to the decreasing of these parameters
due to the expansion. The interplay of such conflicting
tendencies provides non-monotonic behaviour of the $\dot{a}$
function, and the model admits the existence of both periods of
evolution: expansion with acceleration and expansion with
deceleration. The model under consideration requires to
investigate the next step - the estimation of the parameters,
which have been phenomenologically introduced into the force term.
This task is planned to be fulfilled in the next paper
\cite{prepar}.

\vspace{5mm}

\noindent {\bf\Large Acknowledgments}

\vspace{2mm}

\noindent The author is grateful to W. Zimdahl for the fruitful
discussions. This paper was supported by the Russian Foundation
for Basic Research, Russian Program of Support of the Leading
Scientific Schools (grant HW-1789.2003.02), Deutsche
Forschungsgemeinschaft, and NATO (grant PST. CLG.977973).


\begin{thebibliography}{99}

\bibitem{RT} Turner, M. S.,  and Riess, A. Do SNe Ia provide direct
evidence for past deceleration of the Universe?
(astro-ph/0106051).

\bibitem{Riess} Riess, A.G. et al. (1998). {\it Astron. J.} {\bf 116}, 1009;
Schmidt, B. et al. (1998). {\it Astrophys. J.} {\bf 507}, 46;
Perlmutter, S. et al. (1999). {\it Astrophys. J.} {\bf 517}, 565;
Riess, A.G. (astro-ph/0005229); de Bernardis, P. et al. (2000).
{\it Nature} {\bf 404}, 955; Hanany, S. et al. (2000). {\it
Astrophys. J. Lett.} {\bf 545}, 5.

\bibitem{aastar1} Sahni, V., and Starobinsky, A.A. (2000). {\it Int. J. Mod. Phys. D} {\bf 9}, 373.

\bibitem{Ellis} Ellis, J. (2003). {\it Phil.Trans.Roy.Soc.Lond.} {\bf A361}, 2607.

\bibitem{RaPeeb} Ratra, B., and Peebles, P.J.E. (1988).
{\it Phys. Rev. D} {\bf 37}, 3406; Wetterich, C. (1988). {\it
Nucl. Phys. B} {\bf 302}, 668.

\bibitem{caldwell}
Frieman, J.A., Hill, C.T., Stebbins, A., and Waga, I. (1995). {\it
Phys. Rev. Lett.} {\bf 75}, 2077; Caldwell, R.R., Dave, R., and
Steinhardt, P.J. (1998). {\it Phys. Rev. Lett.} {\bf 80}, 1582;
Zlatev, I., Wang, L., and Steinhardt, P.J. (1999). {\it Phys. Rev.
Lett.} {\bf 82}, 986; Faraoni, V. (2000). {\it Phys. Rev. D} {\bf
62}, 023504.

\bibitem{amendola} Amendola, L. (2000). {\it Phys. Rev. D} {\bf 62}, 043511; Amendola, L., and
Tocchini--Valentini, D. (2001). {\it Phys. Rev. D} {\bf 64},
043509.

\bibitem{ZP} Zimdahl, W., Pav\'{o}n, D., and Chimento, L.P.
(2001). {\it Phys. Lett. B} {\bf 521}, 133; Zimdahl, W,. and
Pav\'{o}n, D. (astro-ph/0210484).

\bibitem{Matar}  Baccigalupi, C., Matarrese, S., and Perrota, F.
(2000). {\it Phys. Rev. D} {\bf 62}, 123510.

\bibitem{Uzan} Uzan, J.-Ph. (1999). {\it Phys. Rev. D} {\bf 59}, 123510;
Amendola, L. (1999). {\it Phys. Rev. D} {\bf 60}, 043501;  Chiba,
T. (1999). {\it Phys. Rev. D} {\bf 60}, 083508; de Ritis, R.,
Marino, A.A., Rubano, C., and Scudellaro, P. (2000). {\it Phys.
Rev. D} {\bf 62}, 043506; Boisseau, B., Esposito-Farese, G.,
Polarski, D., and Starobinsky, A.A. (2000). {\it Phys. Rev. Lett.}
{\bf 85}, 2236.

\bibitem{Capo} Capozziello, S. (2002). {\it Int. J. Mod. Phys.D} {\bf 11}, 483.

\bibitem{Frolov} Frolov, A., Kofman, L., and Starobinsky, A. (2002).
{\it Phys. Lett B} {\bf 545}, 8.

\bibitem{ZiBa01} Zimdahl, W., and Balakin, A.B. (2001).
{\it Phys. Rev. D} {\bf 63}, 023507.

\bibitem{ZSBP}  Zimdahl, W., Schwarz, D.J., Balakin, A.B.,  and
Pav\'{o}n, D. (2001). {\it Phys. Rev. D} {\bf 64}, 063501;
Zimdahl, W., Balakin, A.B., Schwarz D.J., and Pav\'{o}n, D.
(2002). {\it Grav. Cosmol.} {\bf 8}(Suppl.II), 158.

\bibitem{ZBEntr} Zimdahl, W., and Balakin, A.B. (2002). {\it
ENTROPY} {\bf 4}, 49. (gr-qc/0109081).

\bibitem{S} Schwarz, D.J. (astro-ph/0209584).

\bibitem{NJP} Balakin, A.B., Pavo'n, D., Schwarz, D.J., and Zimdahl,
W. (2003). {\it New J. Phys.} {\bf 5}, 85.1 - 85.14.
(astro-ph/0302150).

\bibitem{BLL} Laptev, B.L. (1958). In {\it Geometrija i teoriya otnositelnosti},
Kazan University Press, Kazan, p.75 (in Russian).

\bibitem{AAV} Vlasov A.A. (1966). {\it Statistical Distribution Functions},
Nauka, Moscow, (in Russian).

\bibitem{Ehl} Ehlers, J. (1971). In
{\it General Relativity and Cosmology}, Sachs, B.K. (ed.),
Academic Press, New York, pp. 1-70.

\bibitem{Stew} Stewart, J.M. (1971). {\it Non-equilibrium Relativistic
Kinetic Theory}, Springer, New York.

\bibitem{Groot} de Groot, S.R., van Leeuwen, W.A., and van Weert,
Ch. G. (1980). {\it Relativistic Kinetic Theory}, North Holland,
Amsterdam.

\bibitem{RH} Hakim, R. (1968). {\it J. Math. Phys.}{\bf 9}(1), 116-130.

\bibitem{WI} Israel, W. (1978). {\it Gen. Relat. Grav.} {\bf 9}(5), 451-468.

\bibitem{FT} Feldman, Y., and Tauber, G.E. (1980). {\it Gen. Relat. Grav.} {\bf 12}(10), 837-856.

\bibitem{EH} Elze, H.T.,  and Heinz, U. (1989). {\it Phys.Rep.} {\bf 183}, 81.

\bibitem{LI} Litim, D.F., and Manuel, C. (2002). {\it Phys.Rep.} {\bf 364}, 451.

\bibitem{Balescu} Balescu, R. (1975). {\it Equilibrium and Non - Equilibrium Statistical Mechanics},
Wiley, New York.

\bibitem{98} Zimdahl, W., and Balakin, A.B. (1998). {\it Phys.Rev.D}{\bf 58} 063503 (1-10).

\bibitem{mult} Landa, P.S., and  McClintock, P.V.E. (2000). {\it Phys. Rep.} {\bf 323}, 1-80.

\bibitem{prepar} Zimdahl, W., and  Balakin, A.B. (in preparation).

\end{thebibliography}
\end{document}